\documentclass[a4paper,twocolumn]{article}
\date{}
\usepackage{cite}

\usepackage[dvipdfm]{graphicx}
\graphicspath{{./images/}}
\DeclareGraphicsExtensions{.eps}

\usepackage[hyphens]{url}
\usepackage{amssymb}
\usepackage{tcolorbox}
\usepackage{enumitem} 
\usepackage{authblk}

\begin{document}
\title{Users Feel Guilty: Measurement of Illegal Software Installation Guide Videos on YouTube for Malware Distribution}

\author[1,2]{Rei Yamagishi}
\author[1]{Shota Fujii}
\author[2]{Tatsuya Mori} 
\affil[1]{Hitachi, Ltd.} 
\affil[2]{Waseda University} 

\maketitle

\begin{abstract}
This study introduces and examines a sophisticated malware distribution technique that exploits popular video sharing platforms. In this attack, threat actors distribute malware through deceptive content that promises free versions of premium software and game cheats. Throughout this paper, we call this attack {\em MalTube}. 
MalTube is particularly insidious because it exploits the guilt feelings of users for engaging in potentially illegal activity, making them less likely to report the infection or ask for a help. To investigate this emerging threat, we developed video platform exploitation reconnaissance {\em VIPER}, a novel monitoring system designed to detect, monitor, and analyze MalTube activity at scale. Over a four-month data collection period, VIPER processed and analyzed 14,363 videos, 8,671 associated channels, and 1,269 unique fully qualified domain names associated with malware downloads.
Our findings reveal that MalTube attackers primarily target young gamers, using the lure of free software and game cheats as infection vectors. The attackers employ various sophisticated social engineering techniques to maximize user engagement and ensure successful malware propagation. These techniques include the strategic use of platform-specific features such as trending keywords, emoticons, and eye-catching thumbnails. These tactics closely mimic legitimate content creation strategies while providing detailed instructions for malware infection.
Based on our in-depth analysis, we propose a set of robust detection and mitigation strategies that exploit the invariant characteristics of MalTube videos, offering the potential for automated threat detection and prevention.

\end{abstract}

\section{Introduction}\label{sec:introduction}
YouTube~\cite{UnknownUnknown-oi} has become a popular media platform for Internet users worldwide. As of June 2024, youtube[.]com ranked 6th in the Tranco list\footnote{Available at https://tranco-list.eu/list/G62YK.} ~\cite{Le-Pochat2019-nw} of domain access rankings. An estimated 2,504 million active users were reported on the platform in April 2024~\cite{Statista-media-2024}. However, this popularity has also attracted cybercriminals, with reports of scams, malicious URL-embedded comments, and other illicit activities proliferating on the platform~\cite{Bouma-Sims2021-td,Li2024-gy,Alshamrani2020-mu}. The combination of a large user base and the potential for malicious exploitation makes YouTube a critical area for cybersecurity research and intervention.

A particularly concerning trend is the distribution of malware disguised as illegal software through YouTube videos~\cite{Lin2024-uk,Pavan2023-qw}. These videos often purport to offer game cheats or cracking tools for premium software. Users who follow the instructions and download the software from URLs provided in video descriptions or comments inadvertently infect their devices with information-stealing malware. This attack vector, MalTube, has been observed since late 2022. MalTube is particularly insidious because it preys on the guilt feelings of users for engaging in potentially illegal activity, making them less likely to report the infection or seek help. This psychological manipulation adds a layer of complexity to the threat, effectively silencing victims and prolonging the effectiveness of the attack. Despite its persistence and sophistication, comprehensive studies of MalTube are scarce, with existing reports focusing primarily on isolated incidents rather than providing a systematic analysis of the MalTube ecosystem, attacker targeting strategies, and sophisticated user deception techniques.

This phenomenon is consistent with broader trends in cybersecurity incidents: 46\% of corporate incidents are attributed to employee negligence or lack of information, and over 40\% of companies report that employees tend to hide incidents~\cite{Frenkel2017-xe}. Similarly, experts have noted that victims of cyberattacks on pornographic websites often hide their experiences out of shame, a tendency exploited by attackers~\cite{Pilici2020-rj,Doffman2019-hu}. The reluctance to report or deal with infections can lead to prolonged exposure and increased risk for both individual users and organizations. This pattern underscores the need for proactive detection and mitigation strategies that do not rely solely on user reporting.

Given these observations, we aim to conduct a systematic study of the MalTube ecosystem and develop effective countermeasures based on our findings. Specifically, we address the following research questions, each of which is critical for understanding and combating the threat:

\begin{description}
    \item[RQ1:] What user demographics is MalTube targeting?
    \item[RQ2:] What are the characteristics of channels associated with MalTube?
    \item[RQ3:] What are the characteristics of MalTube videos?
    \item[RQ4:] What infrastructure supports the operation of MalTube?
\end{description}

To answer these questions, we designed and implemented video platform exploitation reconnaissance ({\textsc VIPER}), an attack monitoring system. VIPER regularly discovers new MalTube videos, collects associated information such as thumbnails and comments, and continuously monitors these videos to track changes over time. This approach reveals user manipulation techniques embedded in videos and thumbnails, as well as the ephemeral nature of some videos and URLs. 

By analyzing the collected data, we infer attack targets based on the types of fake software and the languages used in the videos (RQ1). This involves identifying the demographics that are most likely to be enticed by the specific types of software being falsely advertised and understanding the linguistic and cultural nuances that make certain populations more vulnerable. We also illuminate the characteristics of MalTube-associated channels (RQ2) by examining patterns in channel creation, activity levels, and social engineering tactics used to build trust and credibility among viewers. Furthermore, we analyze the content, presentation, and engagement metrics of MalTube videos (RQ3), which helps us understand how these videos are optimized to attract and retain viewer attention. Finally, we scrutinize the URLs promoted in these videos to map the infrastructure of attackers (RQ4), revealing the network of web domains, hosting services, and backend systems that support the dissemination and control of the malware. This comprehensive analysis allows us to develop a detailed profile of the MalTube ecosystem, providing critical insights into its operational mechanics and vulnerabilities.

To the best of our knowledge, this study represents the first systematic investigation of the MalTube ecosystem using a dedicated monitoring system. Our key contributions are summarized below.

\begin{itemize}[leftmargin=0.2in]
\item The design and implementation of \textsc{VIPER}, a specialized monitoring system for MalTube. Over a four-month period, this system monitored 14,363 in-the-wild MalTube instances and discovered 1,269 unique fully qualified domain names (FQDNs) used for malware downloads.
\item The primary target demographic was identified as young gamers interested in cheat tools, with a particular focus on Roblox-related content. This highlights a significant risk to children, who are more susceptible to temptation and have limited security knowledge.
\item Analysis of the sophisticated engagement tactics of attackers, including the use of multilingual keywords, emojis, and character images in video descriptions and thumbnails. We also identified distinctive MalTube features, such as repetitive instructions and password-protected archives, that can inform future detection and mitigation strategies.

\item Developed targeted recommendations for platform vendors, security researchers, and end users to effectively combat the MalTube threat.
\end{itemize}

This research provides critical insights into the MalTube phenomenon, laying the groundwork for more effective countermeasures and highlighting the need for targeted education and awareness campaigns to protect vulnerable user groups. By comprehensively mapping the MalTube ecosystem, we enable a more nuanced understanding of this emerging threat and pave way for innovative defenses. Our findings underscore the importance of collaborative efforts between platform providers, security researchers, and end users to mitigate the risks associated with malware distribution through popular video-sharing platforms.
\section{Background}\label{sec:background}
\subsection{Malware Distribution to Users' Computers}\label{subsec:malware_distribution}

Malware is a malicious application that infects the computer of a user and achieves the goals of an attacker. For example, attackers use ransomware, a type of malware, to encrypt the files of victims as a threat to demand money. Attackers also use malware to steal information, such as personal information, authentication information for various services, and information about the company they work for and its internal network, with the aim of performing further attacks. To infect the computer of the user with this type of malware, the attacker first attempts to distribute the malware. The classic distribution method is for attackers to send emails with malware attachments to a large number of users to spread the infection~\cite{PhishLabs2018-tr,UnknownUnknown-pg}. In a drive-by download attack~\cite{Sood2016-ie}, the attacker compromises a website and sets it up to redirect victims to another website. When the victim visits the site, they are repeatedly redirected to multiple sites that act as relays and ultimately deliver malware. In addition, the methods used by attackers to distribute malware are constantly evolving, with methods such as supply chain attacks~\cite{UnknownUnknown-gc} and technical support scams~\cite{Rauti2017-ge}. In this context, MalTube has been observed as a new technique.

\subsection{Overview of MalTube}\label{subsec:Maltube}

\begin{figure}
    \centering
    \includegraphics[width=0.9\linewidth]{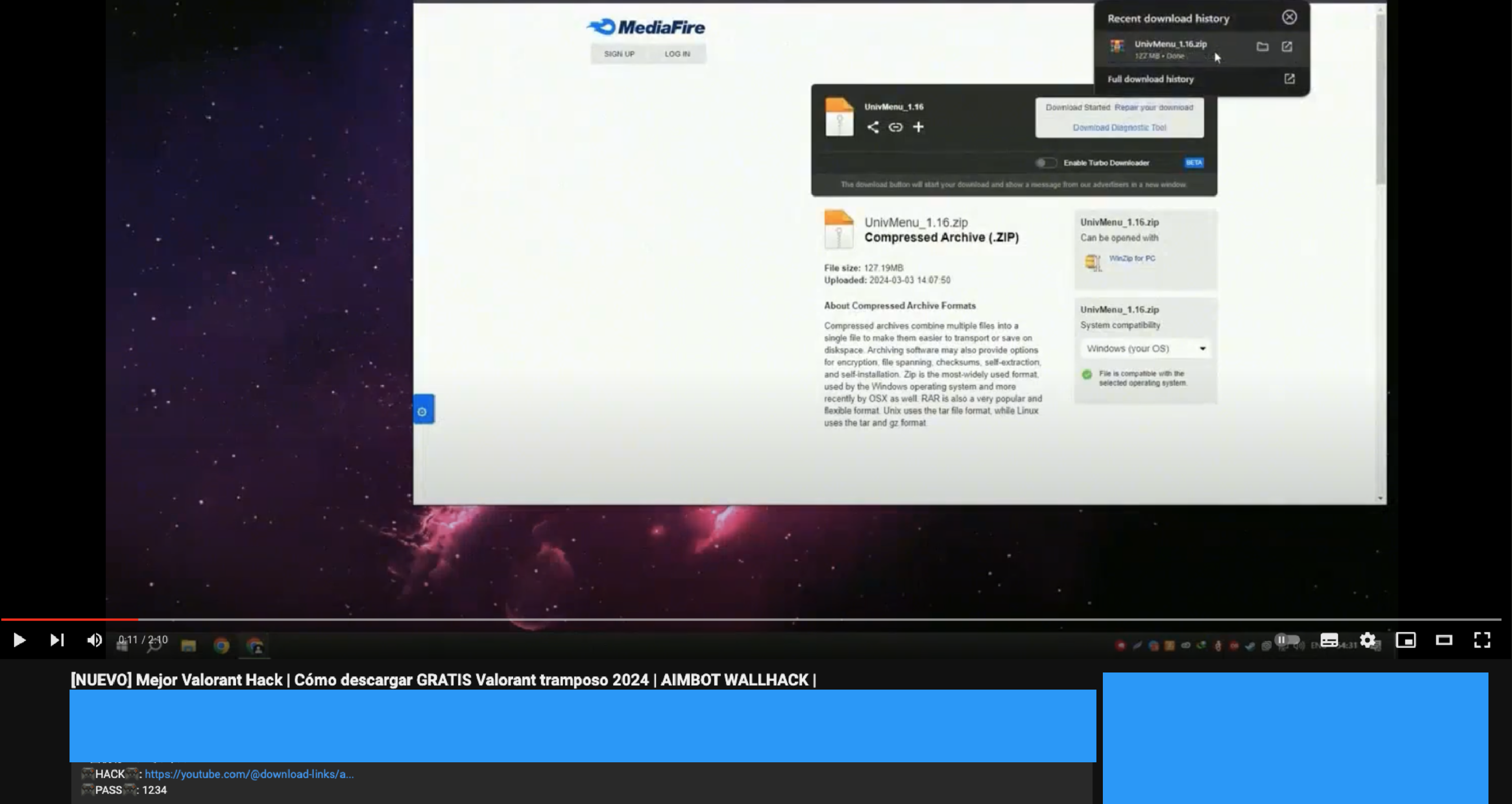}
    \caption{Example of MalTube.}
    \label{fig:example_Maltube}
\end{figure}

YouTube videos about installing of illegal software have been observed as a new way of spreading malware (MalTube). MalTube is reported not in the research paper but in articles~\cite{Lin2024-uk,Pavan2023-qw,Unknown2023-vq}. According to the report, this technique has been observed since late 2022, and the posted video, as shown in Figure~\ref{fig:example_Maltube}, is a video of a game cheat (a tool that provides illicit features such as gaining abilities or increasing ability parameters that other users do not possess) and a prominent software cracking tool.
URLs are included in the description and comments of the videos shown at the bottom of Figure~\ref{fig:example_Maltube}, and users are infected with malware (information theft) by downloading software via the sites directed by the URLs.

Similarly, the articles reported the following characteristics involved in this attack.
\begin{itemize}[leftmargin=0.2in]
    \item Some of the videos posted are examples of AI-generated human speakers to give viewers a sense of familiarity.
    \item Videos include multiple keywords as tags in the description, which is a search engine optimization (SEO) strategy. Some of the cases include keywords for specific countries and languages.
    \item Channels posting videos include popular channels that have been hijacked and less popular channels.
    \item Sites that distribute malware videos include file-sharing services such as MediaFire~\cite{UnknownUnknown-zi} and proprietary sites that let users download malicious files directly. In some cases, the site may be accessed via Reddit or other intermediary sites
    \item The malware distributed is an information-stealing malware with advanced features such as virtual environment detection, such as Lumma and Racoon Stealers.
\end{itemize}

Because these articles report primarily on case studies of attacks, they do not focus on the details of the systematic attack ecosystem or the targeting/user deception tactics of the attacker.

\section{VIPER Framework}\label{sec:methods}

VIPER is a framework designed to understand the main target/user deception techniques of MalTube.
Here, we first provide an overview of VIPER in Section~\ref{subsec:methodoverview}.
Thereafter, each component of VIPER is described in detail in Sections~\ref{subsec:keywords},~\ref{subsec:newmovies},~\ref{subsec:regularlyapi}, and~\ref{subsec:filtering}.

\begin{figure*}[tb]
    \includegraphics[width=1\linewidth]{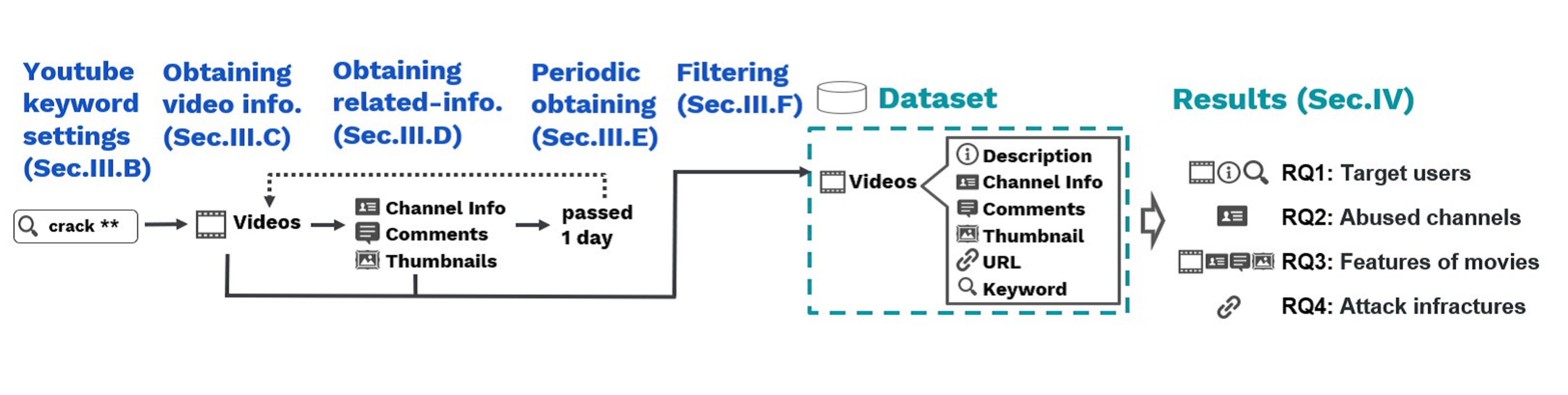}
    \caption{Overview of VIPER.}
    \label{fig:overview}
\end{figure*}

\subsection{The Overview of the VIPER Framework}\label{subsec:methodoverview}

Figure~\ref{fig:overview} shows an overview of VIPER.
VIPER consists of five components: setting keywords, retrieving video information, retrieving related information, retrieving periodic information, and filtering.
VIPER acquires video information on Youtube (Section~\ref{subsec:newmovies}) based on the keywords (Section~\ref{subsec:keywords}) set at the beginning.
Thereafter, additional information related to the video, such as the channel where it was posted, is acquired (Section~\ref{subsec:relatedinfo}).
In addition, regular observation is performed to keep track of updates to the acquired video and channel information (Section~\ref{subsec:regularlyapi}).
Finally, the observed data is filtered (Section~\ref{subsec:filtering}) because the acquired video information contains videos that are not related to the attack to be observed.
The filterd data is our dataset for this study.

\subsection{Keyword Setting}\label{subsec:keywords}

In observing MalTube, it is essential to utilize search keywords that can facilitate the collection of videos posted by the attacker.
In this framework, the keywords are initially defined as input variables to enable the collection of generic observations.

For the observations in this study, to set keywords, several videos were collected as preliminary research and accessed sites that download malware induced by the videos. From the tags of the collected videos and the information on the malware download sites (some malware distribution sites include a list of illegal software that includes other software, such as EC sites), we then increased the variation of keywords. The resultant set of keywords is listed in Table~\ref{tab:keywords}.

\begin{table}
    \centering
    \caption{Keywords Set in VIPER.}
    \scalebox{0.65}{
    \begin{tabular}{|l|l|}\hline 
        Keywords & Related Illegal Software\\\hline 
        adobe+crack & Adobe products \cite{UnknownUnknown-ci}   \\
        avast+premium+security+key & Avast premium security \cite{UnknownUnknown-it}    \\
        Fortnite+cheat & Fortnite \cite{UnknownUnknown-eu}   \\
        internet+download+manager+free & Internet download manager \cite{UnknownUnknown-jk}\\
        Kiddions+cheat & Grand Theft Auto \cite{GamesUnknown-mt} \\
        maxon+redshift+crack & Redshift \\
        microsoft+office+crack & Microsoft Office \cite{UnknownUnknown-tp}  \\
        mw3+unlockall+tool & Call of Duty: Modern Warfare I\hspace{-1.2pt}I\hspace{-1.2pt}I \cite{UnknownUnknown-yh} \\
        roblox+scripts & Roblox \cite{Roblox-CorporationUnknown-xg}   \\
        valorant+cheat & Valorant \cite{UnknownUnknown-sr} \\\hline 
    \end{tabular}}
    \label{tab:keywords}
\end{table}

\subsection{Obtaining Information on Newly Posted Videos}\label{subsec:newmovies}

The VIPER system periodically collects newly posted videos from the search API (/search) of the YouTube Data API v3~\cite{UnknownUnknown-sj}. In order to perform this function, a set of keywords must be input, as well as the search start period and the maximum number of video replies. In this study, the interval for data acquisition was set to one day. Accordingly, the time of the previous acquisition of newly posted video information (approximately one day ago) was selected. Additionally, the maximum number of video replies has been set to 50. The limit on the number of the latest videos that can be retrieved in a single API means that if more than 50 videos are posted in a day, not all of them can be retrieved. Therefore, in this case, the posting period of the 50th video that could be retrieved is set as the search start period, and the videos that could not be retrieved using the Youtube Data API are retrieved again.

Therefore, for each video posted, the video ID, channel ID where the video is posted, title, video posting date, description, specified tags, thumbnail URL, number of views, number of comments, and number of high ratings are returned as responses.
Note that VIPER was implemented in Python3 for the observations in this study, and the Google API Client module~\cite{UnknownUnknown-kf} was used for the Youtube Data API v3.

\subsection{Obtaining Additional Information Related to the Videos}\label{subsec:relatedinfo}

Additional related information such as channel information, comments, and thumbnails, which are not included in each video acquired in Section~\ref{subsec:newmovies}, are obtained. Channel information represents the metadata associated with the channel (account) that posted the video. The aforementioned information is obtained via the /channels API, which encompasses the channel name, channel description, custom URL for the channel's main page, channel launch date, channel country information, number of channel views, number of channel subscriptions, and number of videos posted on the channel. It is possible that comments may contain URLs that direct the user to malicious websites. Therefore, comments made by the channel owner or other users on videos are also retrieved. The comments are obtained with the /comments API. The video ID is specified as a parameter, and a list of comments is returned as a response. Thumbnails are of importance to attackers who wish to attract the attention of as many users as possible, as they provide users with their initial impression. Thumbnails are also obtained by accessing the thumbnail URL in the video search API (/search).

\subsection{Periodic Information Obtaining}\label{subsec:regularlyapi}
In regard to videos, channels, and comments, alterations have been made to the statistical data and information updates. By maintaining an ongoing observation of the statistical information, one can discern changes in the number of views and registered users. Additionally, alterations in the URLs hosted by attackers and the deletion of videos can be identified, which ultimately provides insight into the MalTube ecosystem. In order to address this change, the framework collects information on a periodic basis (at one-day intervals) using the /search, /channels, and /comments API v3.

\subsection{Filtering}\label{subsec:filtering}

The data set included videos of distributors engaged in activities within the game, filmed using game cheats for promotional purposes. These videos were not related to the observed attacks. Consequently, VIPER employed two filtering methods, which were identified through the analysis of the actual data.
\begin{itemize}[leftmargin=0.2in]
    \item Videos that did not include a URL in the descriotions or comments were excluded by filtering because they did not provide paths for malware distribution.
    \item The URLs included in the descriotions or comments were SNS: X(Twitter)~\cite{UnknownUnknown-yr}, Instagram~\cite{UnknownUnknown-km}, and Facebook~\cite{UnknownUnknown-pb}. SNS-only videos were excluded by filtering because they were for the promotional purposes of the distributor.
\end{itemize}

Filtered data is stored as data.
In this observation, the dataset was stored in MongoDB~\cite{UnknownUnknown-vn}.

\section{Measurement Study of MalTube}\label{sec:result}

The data were collected over the period from 2023/12/20 00:00 (UTC+9) to 2024/4/30 00:00 (UTC+9) using the VIPER framework, as detailed in Section~\ref{sec:methods}. As a result, 14,363 videos were collected.

\begin{figure}
    \centering
    \includegraphics[width=1\linewidth]{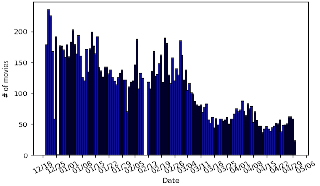}
    \caption{Number of Videos Posted on YouTube in a Day.}
    \label{fig:daily_movies}
\end{figure}

A daily analysis shows that the average number of submissions per day was 110.69 and the median was 120. The maximum number of postings was 252 on 21st Dec (UTC+9).
Figure~\ref{fig:daily_movies} shows the number of daily video postings.
Approximately 150 postings were observed in December, January, and February, although the number of postings varied.
The number of videos posted has decreased since March, suggesting that activity may have decreased or that videos are being deleted after less than a day.
We also observed zero postings on 26th Dec. (UTC+9), 10th Feb. (UTC+9), and 11th Feb. (UTC+9), suggesting that the attacker may have been inactive.

Henceforth, we discuss the results of our data analysis for each RQ. Finally, we evaluate the implications and threat assessment.

\subsection{RQ1:Target Users}\label{subsec:result_rq1}

Section~\ref{subsec:result_rq1} identifies the characteristics of users targeted by MalTube. Because MalTube requires the attacker to make the target view the video, the attacker needs to devise a way to create a video that will attract the interest of the target. Therefore, it is possible to infer the target user demographics based on the topic of the video and the illegal software. Similarly, because we believe that understanding the users who use the target country and language lead to specific countermeasures, we identify the language used in the video.

\subsubsection*{\textbf{Posted videos by keyword and software}}

As indicated in Section~\ref{subsec:keywords}, VIPER retrieves videos based on keywords corresponding to illegal software.
Therefore, it is possible to identify the illegal software that was the topic of the video based on the search keywords that the video is associated with. For example, a video found by searching for ``Fortnite+cheat'' is determined to be a video about Fortnite. Note that because attackers devise ways to ensure that videos are viewed when targets perform keyword searches for the software in question, we believe that focusing on the association between keywords and illegal software in videos is appropriate. Table~\ref{tab:disguised_software} summarizes the number of videos per relevant software and the type of software in question.

\begin{table}
    \centering
    \caption{The Type and Number of Video Topics (Illegal Software).}
    \scalebox{0.65}{
    \begin{tabular}{|l|l|r|}\hline 
       Related Illegal Software  & Categories & Videos\\\hline 
       Roblox \cite{Roblox-CorporationUnknown-xg}  & game platform & 4,305\\
       Valorant \cite{UnknownUnknown-sr} & game & 2,865\\
       Fortnite \cite{UnknownUnknown-eu}  & game & 2,291\\
       Grand Theft Auto \cite{GamesUnknown-mt} & game & 1,884\\
       Call of Duty: Modern Warfare I\hspace{-1.2pt}I\hspace{-1.2pt}I \cite{UnknownUnknown-yh} & game & 1,419\\
       Adobe Premiere Pro \cite{UnknownUnknown-ci} & Movie  & 951\\
       Internet Download Manager \cite{UnknownUnknown-jk} & Download Manager & 122\\
       Avast Software \cite{UnknownUnknown-it} & Security product & 75\\
       Microsoft Office \cite{UnknownUnknown-tp} & Bussiness Software & 25\\\hline 
    \end{tabular}}
    \label{tab:disguised_software}
\end{table}

The software associated with the videos often posed as free access to paid content such as game cheat tools or costumes, and the top five types of videos were related to games (game platforms). Non-game videos frequently presented themselves as installers of Adobe Premiere Pro, a paid software, and claimed to be available for free. 

To obtain more detailed information about MalTube's main target, we analyze the target users by investigating information and user demographics for the top three games. Roblox, which was the most frequently used keyword, is a game platform that allows users to create and share games in a virtual space where they can create and interact with avatars. According to the article~\cite{Dean2023-mk}, as of 3Q 2023, it had 70.2 million active users, with 22.4\% of users in the US and Canada, 26.9\% in Europe, and 23.1\% in Asia-Pacific, showing that its users are spread across the world. In addition, the age group of users is skewed towards the younger generation, with 22\% of users being under nine years, 23\% being 9--12 years, 15\% being 13--16 years, and 21\% being 17--24 years.

Valorant is a tactical shooting game that emphasizes competitiveness. According to the article~\cite{Jarvis-the-NPC2024-qo,Perez2023-kg}, it is estimated that the number of active players exceeded 20 million as of October 2023. , with 23.53\% of users in the US, followed by Brazil at 6.72\%, Turkey at 6.12\%, Russia at 5.95\%, and South Korea at 4.84\%. Although no official data on age groups has been released, the article~\cite{Jarvis-the-NPC2024-qo} analyzed that Valorant is skewed towards younger users, but not as much as other games such as Fortnite, or CSGO (Counter-Strike: Global Offensive), which is skewed towards older users.

Fortnite is a shooting game with crafting elements. According to the article~\cite{Smith2024-db}, as of 2023, 231 million people were active users, and it was being played worldwide. Among them, 62.7\% were aged 18--24, 22.5\% were aged 25--34, and 12.7\% were aged 35—44; however, the article~\cite{Smith2024-db} points out that users under the age of 18 are playing by falsifying their age.

From these trends, it can be inferred that the number of users of software related to videos is large, and that attackers are adopting a strategy of disguising themselves as cheat tools for popular games to attract the interest of many users and spread malware.
In addition, there is little regional bias among game users, and it can be said that the target is users from all over the world.
Conversely, the age distribution of game users is skewed towards younger age groups such as teenagers and 20--29 years (which is perhaps to be expected, given that the target audience is game users).
In particular, the number of video posts related to games like Roblox, which are played by most users under the age of 12, is high, and it has become clear that this includes children who have difficulty making judgments about right and wrong.

\subsubsection*{\textbf{Language of video description}}

For attackers to target users of a particular language, they need to include that language in the video title and description and adopt a strategy that supports video searches in that language. Therefore, to determine the language of the target user, it is necessary to determine the language contained in the video title and description.

We used Fasttext's language identification~\cite{joulin2016bag} (fasttext--langdetect 1.0.5~\cite{UnknownUnknown-in}) to determine the language of the connection of the video titles and descriptions. For videos that did not have description, we only performed language detection on the video title. The results are summarized in Table~\ref{tab:movie_title_lang}.

\begin{table}
    \centering
    \caption{Number of Videos on Language of Titles.}
    \scalebox{0.75}{
    \begin{tabular}{|ll|r|}\hline 
       Language & Codes & \# of videos\\\hline 
       English & en & 13,604\\
        Portuguese & pt & 311\\
        Spanish & es & 162\\
        Russian & ru & 55\\
        German & de & 47\\
        French & fr & 41\\
        Arabic & ar & 39\\
        Others (Turkish, Italian, and so on) & -- & 104\\\hline 
    \end{tabular}}
    \label{tab:movie_title_lang}
\end{table}

English was by far the most popular, with 13,604 videos. There were runner-up videos in Portuguese (311 videos) and Spanish (162). In addition to these, there were 29 different languages, including Turkish, Italian, Japanese, and Indonesian. However, most of the videos were in English only (with no audio, just background music); additionally, we found videos in Portuguese and Spanish.

The fact that most of the videos are in English, a language that is widely used, is consistent with the strategy of attackers who target popular games that are widely used. In addition, there were also videos in English and other languages that had similar characteristics (such as the content of the description); however, there were also videos in Portuguese and Spanish that tended to be different. For example, in English, the audio is just music and there is no human voice, but in other languages such as Portuguese and French, it is possible to see a human speaking and explaining in the relevant language. From these points, it is possible that the attackers who post videos in Portuguese and Spanish are different from the attackers who post videos in English, suggesting that videos are being posted by multiple attackers.

\begin{tcolorbox}
\textbf{Takeaway1}\\
MalTube mainly distributed malware disguised as cheat tools for popular games around the world.
Therefore, it is targeting a wide range of game users, including minors.
In addition, most of the videos posted were in English, which is consistent with the strategy of targeting a wide range of users.
\end{tcolorbox}

\subsection{RQ2:Channels Related to MalTube}\label{subsec:result_rq２}

The 14,363 videos were posted by 8,671 channels.
The statistical information for the channels is summarized in Table~\ref{tab:account_statics}.

In the section below, we analyze these channels.

\begin{table}
    \centering
    \caption{Statistics of the MalTube (YouTube Channels) .}
    \scalebox{0.65}{
    \begin{tabular}{|l|rrr|rr|}\hline 
         & Mean & Mdn. & Var. & Max. & Min. \\\hline 
         Views & 1.30\(\times 10^{6}\) & 274 & 1.57  & 4.90\(\times 10^{8}\) & 0 \\
        Subscribers & 8,367.5 & 18 & 5,02\(\times 10^{9}\) & 3.57\(\times 10^{6}\) & 0\\
        Video Uploads & 189.0 & 10 & 1.28\(\times 10^{8}\) & 1.05\(\times 10^{6}\)& 1\\\hline 
    \end{tabular}}
    \label{tab:account_statics}
\end{table}

\subsubsection*{\textbf{Distribution of the number of video submissions on channels}}

\begin{figure}
    \centering
    \includegraphics[width=1\linewidth]{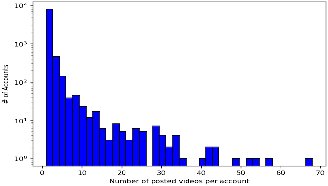}
    \caption{Distribution of the Number of Posted Videos Per Channels.}
    \label{fig:video_per_account}
\end{figure}

The distribution of the number of video posts on channels is shown in Fig.~\ref{fig:video_per_account}.
The channels with the most video posts had 68 videos, followed by 57 videos and 53 videos.
Conversely, 77.7\% of channels had one post, 90.7\% had three or fewer posts, and 94.54\% had three or fewer posts.
Although there are also channels that have posted multiple videos, 98.7\% of channels had posted less than ten videos, indicating that there is a tendency for channels to post fewer videos.

\subsubsection*{\textbf{Distribution of time since registration of channels information}}

\begin{figure}
    \centering
    \includegraphics[width=1\linewidth]{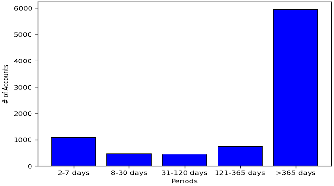}
    \caption{Distribution of Time since Registration of Channels.}
    \label{fig:time_since_registration}
\end{figure}

Figure~\ref{fig:time_since_registration} shows the time between registration and the first video related to an attack being posted. 68.5\% of the channels were created more than 365 days ago, followed by 12.5\% of the channels created 2--7 days ago. Most channels were created more than a year ago, but there was also a trend for many channels to have been created recently. We discuss the channels created more than a year ago below. There were also channels that continued to post only videos related to MalTube; additionally, there were channels that posted videos in different languages regardless of MalTube (Details of the case are introduced at the end of Section~\ref{subsec:result_rq２}). From observations of these channels, it is possible that the attacker is not creating channels for MalTube purposes each time. The report article on this attack~\cite{Lin2024-uk} established that YouTube channels of infected PCs are stolen and misused for video posting. Therefore, it is possible that the channels created more than a year ago were also stolen from victims by the attacker.

Conversely, the emergence of new channels was also observed. It is suspected that these channels were newly created by the attacker with the aim of avoiding detection by YouTube and spreading the risk when disabling.
If MalTube is run on a specific channel, the number of unauthorized activities on that channel will increase, making it easier for YouTube to detect.
In addition, if a channel is disabled owing to detection, the videos it has posted will also be deleted; therefore, it is advisable to distribute the channels that post videos.
Therefore, it is assumed that attackers are creating new channels.

\subsubsection*{\textbf{Country information for channels}}
\begin{figure}
    \centering
    \includegraphics[width=1\linewidth]{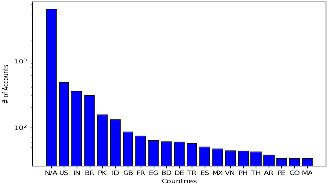}
    \caption{Countries Registered in Channels.}
    \label{fig:countries_account}
\end{figure}

You can optionally set country information for YouTube channels.
Figure~\ref{fig:countries_account} shows the number of channels for each country registered on channels.
Five thousand, nine hundred seventeen channels did not have any country information set (N/A).
The breakdown of the countries with country information set is as follows: the United States (472), India (352), Brazil (299), Pakistan (154), Indonesia (131), the United Kingdom (85),
France (74), Egypt (63), and Bangladesh (63).
In total, 103 different countries were set, including other countries.

\begin{figure}
    \centering
    \includegraphics[width=0.8\linewidth]{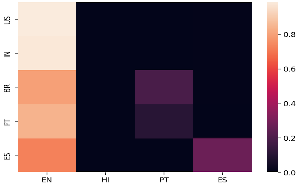}
    \caption{Relationship between the Countries registered in Channels and Language in Video Desctiptions.}
    \label{fig:countries_account_and_lang}
\end{figure}

There were 103 different countries in the channels, but the languages of the titles and descriptions of the videos mentioned in Section~\ref{subsec:result_rq1} were 29 different languages.
There are countries that speak the same language; therefore, it is reasonable that there are many countries in the channels, but this gap is large.
This raises the question of the relationship between the countries of the channels and the languages of the titles and descriptions of the videos.
It is difficult to list them all; Fig.~\ref{fig:countries_account_and_lang} shows the correspondence between the countries and languages that ranked highest in each category using a heat map.
The vertical axis shows the United States (English), India (English), Brazil (Portuguese), Portugal (Portuguese), and Spain (Spanish); the horizontal axis shows the languages English, Hindi, Portuguese, and Spanish.
The languages in parentheses after the countries mentioned above are official languages.

From the heat map, we can observe that English is consistently used.
This is reasonable because English accounts for 94.7\% of all videos (13,604 videos).
In addition, there are videos posted in Brazil, Portugal, and Spain in the official language and in other languages, but most of them were in English.
Interestingly, Indian channels were observed next to American channels; however, it was discovered that they were posting in English, not Hindi. There was only one Hindi-language video, and the country information for the channels was not set.

The gap between the country settings and the languages used in these channels suggests that the channels have been hijacked.
For example, one channels with India's country information set up posted 57 videos related to travel and other topics in Hindi from 2020 to November 2023, and it was a channel with more than 74,000 subscribers.
From December, seven videos related to this campaign (videos introducing cheat tools for GTA5, Valorant, Roblox, and Fortnite, as well as illegal Adobe Premiere Pro installations) were posted, and these videos were in English.
The change in language and video topic trends suggests that the channels may have been hijacked or bought and sold.

\begin{tcolorbox}
\textbf{Takeaway2}\\
The most common case is where more than a year has passed since the channel was created, and the next most common case is where it was created in a short period of time.
Based on an analysis of the language and video content, it is highly likely that the channels that have been around for over a year are owned by different people, and that there has been hijacking or trading.
\end{tcolorbox}

\subsection{RQ3:Video Features}\label{subsec:result_rq3}

The number of views, positive ratings, and comments as of April 30th, 2024, or immediately before the video was deleted, were tallied and are compiled in Table~\ref{tab:movie_statics}.
The variance and average for both views, positive ratings, and comments are high, whereas the median is low,
there are videos that are attracting attention, such as outliers.
The low median value indicates that overall, there are many videos with low view counts, low ratings, and few comments, indicating that they are not very popular.

\begin{table}
    \centering
    \caption{Statistics of the MalTube (Videos) .}
    \scalebox{0.8}{
    \begin{tabular}{|l|rrr|rr|}\hline 
         & Mean & Mdn. & Var. & Max. & Min. \\\hline 
        Views & 1,517.2 & 48 & 9.76\(\times 10^{7}\) & 8.59\(\times 10^{5}\) & 0 \\
        Favorites & 51.84 & 2 & 9.80 \(\times 10^{4}\)& 25,665 & 0\\
        Comments & 14.60 & 1 & 2,954.3 & 2,053 & 0\\\hline 
    \end{tabular}}
    \label{tab:movie_statics}
\end{table}

\subsubsection*{\textbf{Video existence period}}

\begin{figure}
    \centering
    \includegraphics[width=1\linewidth]{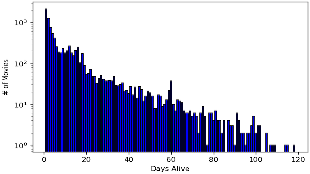}
    \caption{Period Until the Video Deletion.}
    \label{fig:period_deletion}
\end{figure}

Figure~\ref{fig:period_deletion} shows the period between when a video is observed and when it is deleted.
Of the 14,363 videos deleted between December 20, 2023 and April 30, 2024, 5,208 were deleted within a short period.
Many videos were deleted within a short period: 2,162 were deleted in one day, 1,241 were deleted in two days, and 758 were deleted in three days.
From the above, it can be observed that while there are videos that are not deleted for long periods, many videos are deleted within a short period (1--3 days).

\subsubsection*{\textbf{Video description}}
\begin{figure}
    \centering
    \includegraphics[width=0.8\linewidth]{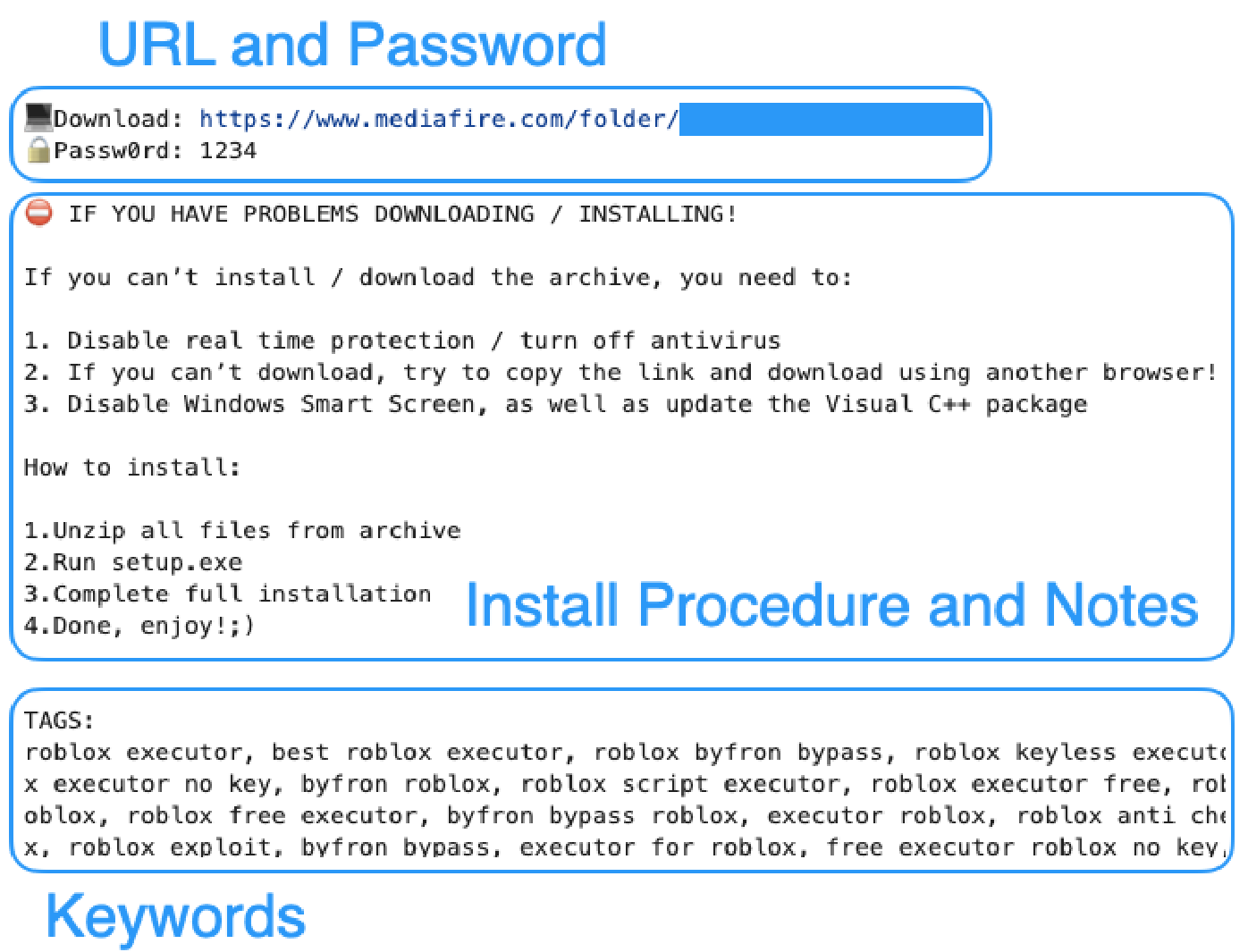}
    \caption{Case of Video Description in MalTube.}
    \label{fig:description_case1}
\end{figure}

The text in the description contained word and structure features.
The most common case, as shown in Fig.~\ref{fig:description_case1}, was found in 6,794 cases.
It consisted of a download (relay) URL, a password for extracting the malware archive, a description (notes and instructions), tags, and keywords.
Because malware is in the form of a password-protected zip file for the purpose of evading detection, it is necessary to communicate the password to the user.
Therefore, words such as ,  ``Password,'' ``Pass,'' ``Pw,''  their leet and emoji of ``Key'' were included in the description.
This password is also repeatedly included in thumbnails, videos, comments, and intermediate sites, and is currently one of the features that identifies this MalTube.
It is assumed that they are trying to avoid detection by using pictograms and reht, but because it is also necessary to correctly convey the password to the target user,
communicating this password could be causing the attackers problems.
Furthermore, the instructions (notes and procedures) were visually explained in the video, but they were also repeatedly explained in the overview section; therefore, users could download without confusion.

In addition, as shown in Fig.~\ref{fig:description_case1}, there was a tendency to use a lot of pictograms. There are also research results showing that pictograms promote purchasing in the advertising field~\cite{Li2022-ah}, so it can be inferred that pictograms are used to give users a sense of familiarity and to promote installation. Moreover, there were also cases where texts similar to this example were included in the comments section rather than descriptions.

\subsubsection*{\textbf{Video thumbnails}}

\begin{table*}
    \centering
    \caption{Features of Raomdome-Picked Thumbnails about Videos.}
    \scalebox{0.7}{
    \begin{tabular}{|l|rrrrrrrr|}\hline 
         &Game Characters& Other Game Characters & Anime Characters & Software Screen & How to Download & Password & Text & Logo\\\hline 
         Roblox& 1/10 & 1/10 & 4/10 & 8/10 & 4/10 & 0/10  & 10/10& 0/10\\
         Valorant & 6/10 & 0/10 & 1/10 & 7/10 & 5/10 & 2/10  & 7/10& 0/10\\
         Adobe& -- & 0/10 & 3/10 & 3/10 & 1/10 &  0/10 & 7/10& 7/10\\\hline 
    \end{tabular}}
    \label{tab:features_thumbnail}
\end{table*}

In the following, we will analyze the thumbnails used in videos targeting specific software or games.  Thumbnails are one of the factors that gives users their first impression of the video. Research has pointed out the impact of thumbnails and titles on the number of views~\cite{Jang2024-zt,Hoiles2017-ah} and the impact of the impression of thumbnails on advertisements in videos~\cite{Hoiles2017-ah}. A random sample of 10 videos targeting Roblox, Valorant (the game with the highest number of videos), and Adobe Premiere Pro (the non-game video with the highest number of views) was selected for manual analysis of the corresponding thumbnails. In the analysis, thumbnails were classified according to the following criteria: whether they depicted game or anime characters, whether they displayed the game or software interface after installation, whether they illustrated the download method and passwords, and whether they included text such as titles or logos. The results of this analysis are presented in Table~\ref{tab:features_thumbnail}.

As a result, when comparing games (Roblox, Valorant) and Adobe Premiere Pro, games tended to have more elaborate thumbnails.
Game video thumbnails often include the screen or characters after installation, and it is easy to attract the interest of users by posting the effects of cheats or friendly characters.
Conversely, in Adobe Premiere Pro, product logos often appeared instead of characters, and the creators were trying to make an impression on users who saw the thumbnails.
In addition, even for videos of the same game, Valorant used characters from the relevant game, whereas Roblox used anime and game characters that had nothing to do with the game.
Roblox is a game where users create their own avatars, whereas Valorant is a game where users control a fixed character; therefore, it is easier to present characters that will attract the interest of users.
We think that anime characters are chosen because the main users of Roblox are young people.

No identical thumbnails were found within the sampled range.
Even across the whole set, there were some common parts such as logos and characters on the Thumbnails; however, there was a tendency for the use of completely identical Thumbnails to be rare.
Creating thumbnails is more difficult than creating text; nevertheless, attackers are willing to pay the cost of creating multiple patterns of thumbnails.

\begin{tcolorbox}
\textbf{Takeaway3}\\
In terms of the characteristics of the video, in addition to being able to perform keyword searches using words such as ``cheat,'' the description contained the word ``Password.''
We believe that it is possible to use such specific keywords for identification and detection.
Thumbnails could also be used for detection, but compared to the description, they were more difficult to handle as image data; additionally, there were a lot of variations.
\end{tcolorbox}

\subsection{RQ4: Attack Infrastructure}\label{subsec:result_rq4}

\begin{figure}
    \centering
    \includegraphics[width=1\linewidth]{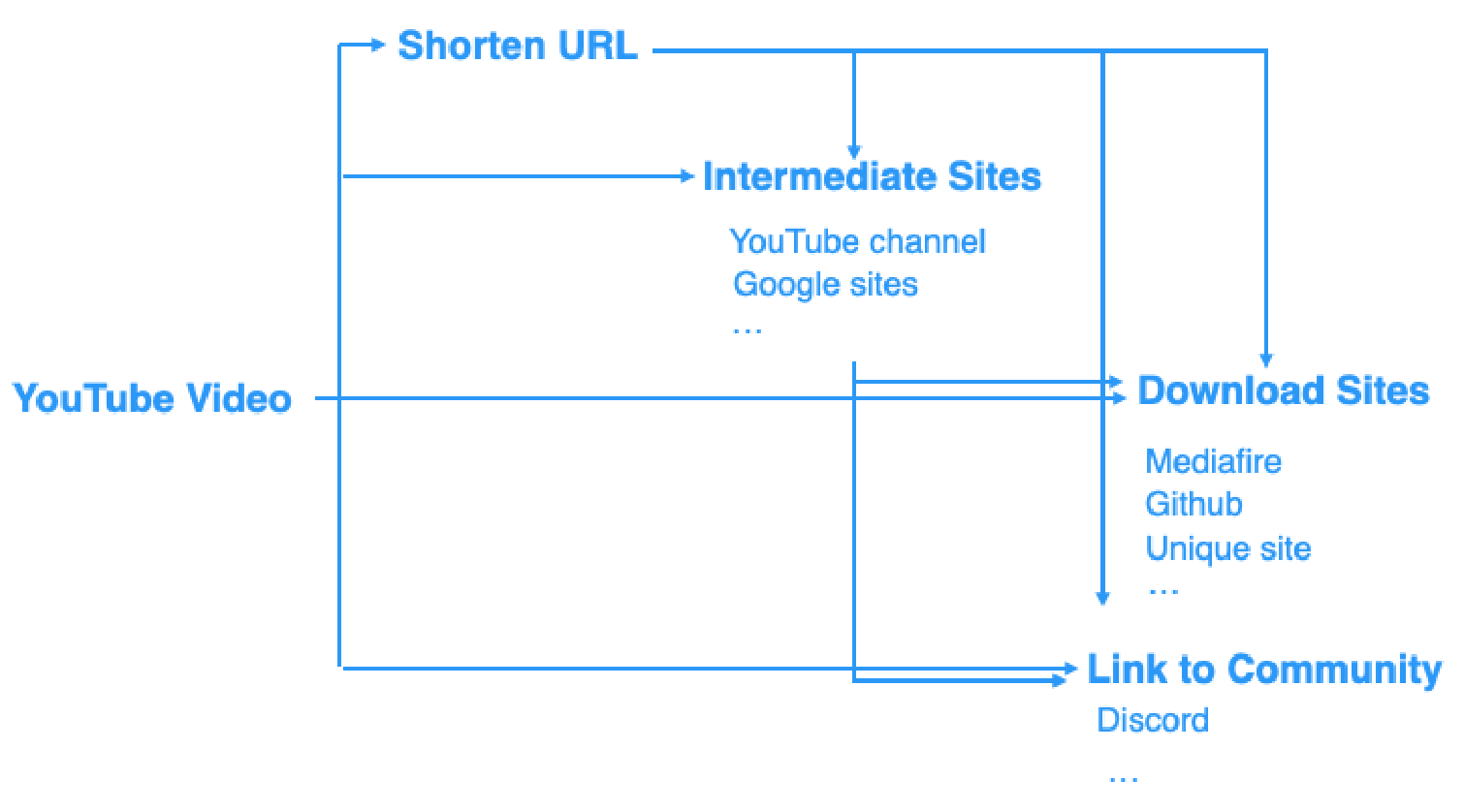}
    \caption{Flow of Links to Malware Download Sites.}
    \label{fig:flow_links}
\end{figure}

Figure~\ref{fig:flow_links} shows the main flow from when a user downloads a video from a site using MalTube.
The URLs in the video direct users to intermediate sites, download sites, and communities.
The download site is the site where the user downloads the malware, and there were cases where the site of attackers was used, as well as cases where a file sharing service was used.
In this study, it was not possible to determine whether all the files downloaded from the download sites were malicious; nevertheless,
we downloaded some of the files and confirmed that they were stealer using online sandboxes and reputation sites (VirusTotal~\cite{UnknownUnknown-av}, Any.run~\cite{UnknownUnknown-sm}, Tria.ge~\cite{UnknownUnknown-oy}).
The term "community" refers to a suspected community where attackers and users communicate directly, such as on discord channels, to encourage malware downloads.
Intermediate sites are those that direct users to download sites.
The role of intermediate sites is to prevent security experts from tracking them, in the same way as drive-by downloads~\cite{UnknownUnknown-zt}.
Conversely, as a difference from drive-by downloads, because the target user has the motivation to download the file,
there is no need for automatic screen transitions on the intermediate site; instead, the user needs to be manually guided to follow the URL.
In addition, within the scope of our observations, the maximum number of relay hops between intermediate sites was two.
This is because it is difficult for the target user to click on multiple URLs in succession, and there is a concern that they will give up downloading the malware.
In addition, shortened URLs were also used, which replaced the URLs of these sites with a shortened format.

\subsubsection*{\textbf{Domain names of suspicious URLs}}

From the description and comments, 31,772 URLs and 1,269 FQDNs were extracted using regular expressions.
This also included URLs that direct users to the official websites of games and other products, as well as promotional social networking sites.
The top and most distinctive domain names are listed in Table~\ref{tab:malicious_domain}.

\begin{table}
    \centering
    \caption{Type of Extracted FQDNs.}
    \scalebox{0.75}{
    \begin{tabular}{|l|l|r|}\hline
        Estimated role& FQDNs & Numbers \\\hline 
        Intermediate site&www[.]youtube[.]com & 5,359\\
        Shortened URL&discord[.]gg & 3,650\\
        Shortened URL&bit[.]ly & 3,385\\
        Download Site&www[.]mediafire[.]com & 2,785\\
       Intermediate site&youtu[.]be  & 1,305\\
       Intermediate site&sites[.]google[.]com  & 896 \\
       Community&discord[.]com  &  661\\
       ...&... & ... \\
       Shortened URL&tinyurl[.]com & 344\\
       Download Site&raw[.]githubusercontent[.]com   & 156\\
       Download Site&telegra[.]ph  & 108\\\hline
    \end{tabular}}
    \label{tab:malicious_domain}
\end{table}

MediaFire\cite{UnknownUnknown-zi} is a file storage service.
MalTube was used as a download site for attackers to distribute malware.
In some cases, after jumping to the URL of another site, the user was directed to download from mediafire[.]com on that site (intermediate site).
In addition, other domains such as drive[.]google[.]com (Google Site~\cite{WorkspaceUnknown-hg}), raw[.]githubusercontent[.]com (GitHub~\cite{UnknownUnknown-cg}), telegra[.]ph (Telegraph~\cite{UnknownUnknown-sg}), genesis-chairs[.]gitbook[.]io, and ibf[.]tw were used to directly direct users to malware download sites.

\begin{figure}
    \centering
    \includegraphics[width=1\linewidth]{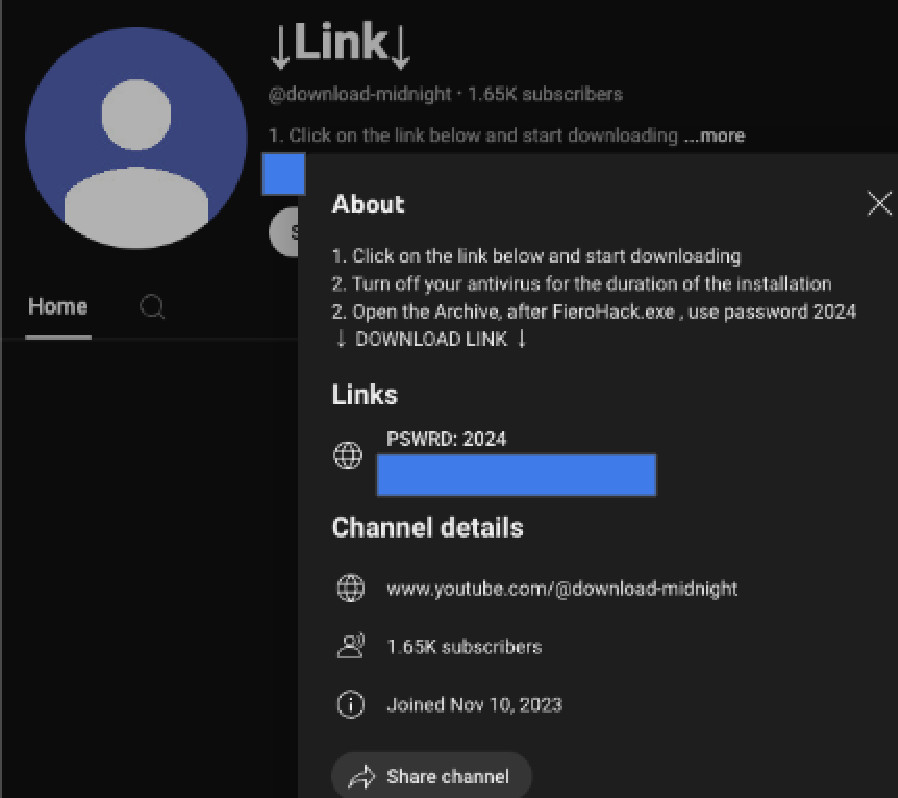}
    \caption{Example of Intermediate YouTube Channel.}
    \label{fig:intermediate_channel}
\end{figure}

\begin{figure}
    \centering
    \includegraphics[width=0.8\linewidth]{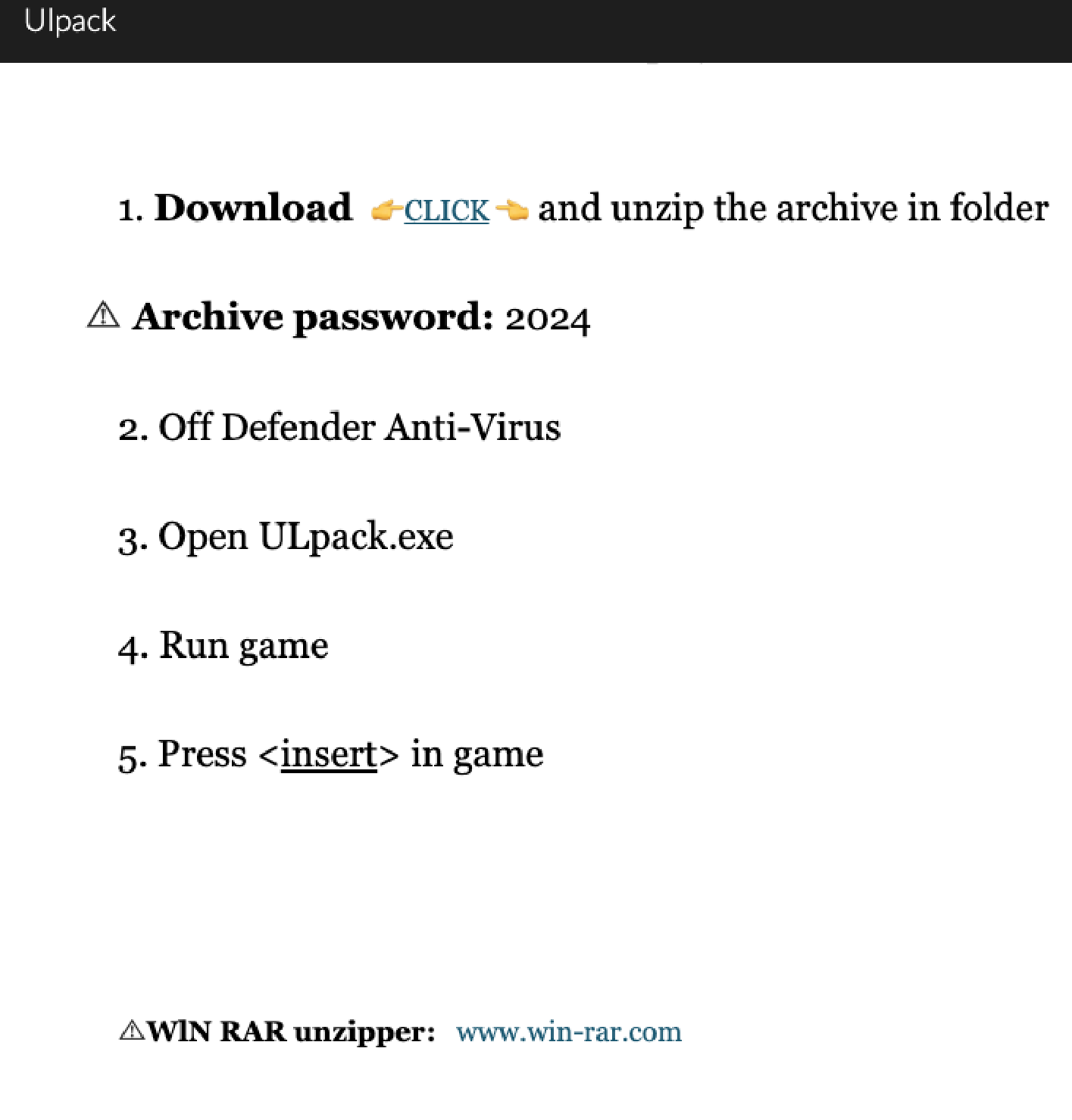}
    \caption{Example of site[.]google[.]com Page.}
    \label{fig:example_google_site}
\end{figure}

Many of the URLs for youtube[.]com and youtu[.]be refer to other Youtube~\cite{UnknownUnknown-oi} channels, and the channel descriptions (Figure~\ref{fig:intermediate_channel}) include text directing users to other sites using the relevant channel as a gateway site.
There were also cases where discord[.]com (Discord~\cite{UnknownUnknown-dc}) directed users to specific communities.
In addition, site[.]google[.]com is also an intermediate site, and it guides users' clicks on Google sites, as shown in Fig.~\ref{fig:example_google_site}.
There is a wide range of sites, from simple sites like Fig.~\ref{fig:example_google_site} to sites that include images that illustrate the effects of the game as well as thumbnails and videos to arouse the interest of users.
Conversely, even simple sites included features that would not cause problems for the target user, such as highlighting the link to click on and restating the download procedure.

In addition, shortened URLs such as bit[.]ly and tinyurl[.]com, as well as URLs that act as proxies for URLs such as discord[.]gg, were included to hide the URL of the attacker.

\begin{figure}
    \centering
    \includegraphics[width=1\linewidth]{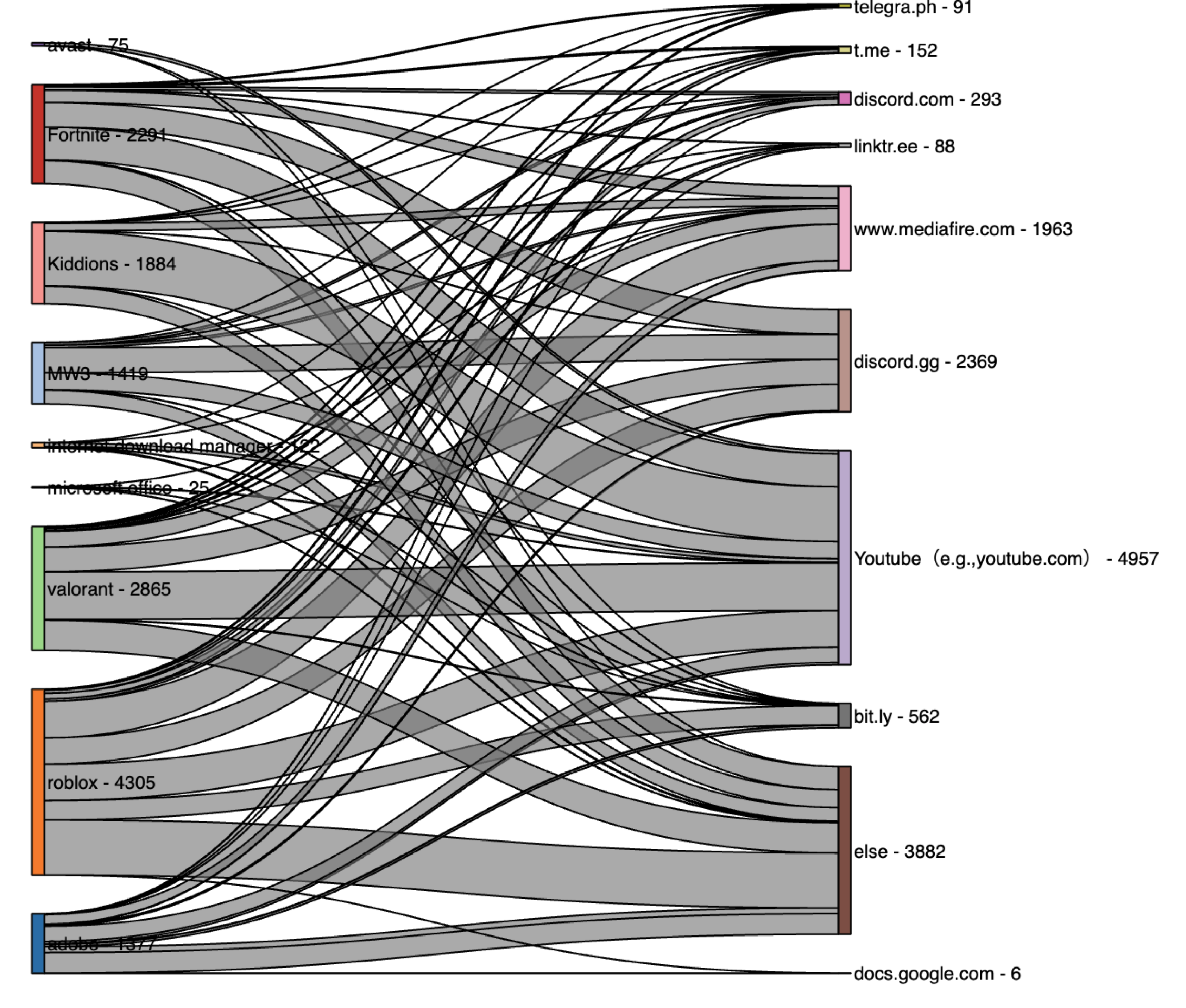}
    \caption{Relation between Video Topic and the Type of Malicious URL.}
    \label{fig:sankey}
\end{figure}

To clarify the relationship between the topics of the videos and the suspicious URLs they lead to, we analyzed what sites were being transitioned to from each target video. The results are shown in Fig.~\ref{fig:sankey}. The relationship between the topics of the videos and the main types of suspicious URLs was found to be insignificant. Furthermore, while there were differences in popularity depending on the type of suspicious URL, they were used in a relatively equal manner. It was anticipated that attack infrastructure and sites specialising in specific game cheat tools, etc., would be prepared to attract users' interest. In some cases, intermediate sites were dedicated to specific games, but within the scope of our observations, there were not many cases that were dedicated to the topic of the video. Our expectations were also contradicted by the fact that the download sites were using general services like mediafire[.]com or marketplaces.

As shown in Fig.~\ref{fig:intermediate_channel}, YouTube was used as intermediate sites. Table~\ref{tab:num_youtube_channel} shows the number of YouTube channel URLs in the video for each channel. In addition to the redirect link and download links, which are relatively straightforward for intermediate sites, there were also distinctive channels, such as TautaraHack and pengwincheat. In certain instances, these distinctive channels resulted in the establishment of particular distribution platforms, such as tauratahack[.]fun and pengwincheat[.]fun. The pertinent websites and online platforms each feature a gecko and penguin character as their logo. Furthermore, the distribution site employs a market-style interface analogous to that of an e-commerce site, offering a variety of game cheat scripts for sale. This system of branding sites with unique characters and directing users constitutes a component of the malware distribution infrastructure.

During the observation period, the URL displayed on the intermediate site's YouTube channel was occasionally modified. Such alterations result in a change of the site to which users are directed. It seems reasonable to posit that this alteration in the destination of the redirect is intended to circumvent analysis, such as drive-by downloads~\cite{UnknownUnknown-zt}.

\begin{tcolorbox}
\textbf{Takeaway4}\\
We confirmed that in some cases, users are directed directly to download sites via URLs in videos and to intermediate sites.
We identified cases where YouTube channels were used as intermediate sites.
Intermediate sites had characteristics that were based on the premise of user clicks, such as the fact that there were few multi-stage relays and that explanations of the download procedure were necessary.
It was also revealed that download sites and intermediate sites make efforts to arouse user interest, such as branding using site-specific character logos.
\end{tcolorbox}

\subsection{Implications and Threat Assessment}\label{subsec:implications}

Our study demonstrates that MalTube poses a significant and diverse threat to a wide range of users, with focus on young gamers. While the attack vector includes various types of illegal software, we observed a preponderance of videos related to game cheating tools. This finding is particularly concerning given the demographics of the target audience.

A notable discovery in our research was the high frequency of videos targeting games with predominantly young user bases, such as Roblox. These videos often feature anime characters or other visually appealing elements in their thumbnails, likely designed to attract the attention of younger viewers. The targeting of minors is particularly alarming considering previous research. Studies have shown that children are more susceptible to temptation owing to their limited moral experience~\cite{Markiewicz2020-at} and often lack sufficient knowledge of cybersecurity best practices\cite{Tirumala2016-wh}. Therefore, children may be more likely than adults to attempt to download game cheating tools and inadvertently infect their devices with malware.

\subsubsection*{\textbf{Guidance via the YouTube Channel}}

\begin{table}
    \centering
    \caption{The Number of the Intermediate YouTube Channel in Videos.}
    \scalebox{0.75}{
    \begin{tabular}{|l|r|}\hline
        Channel Name & Number of Videos to the Channels \\\hline
        @download-midnight & 2,473 \\
        @redirect-link-re8pp & 817\\
        @Download-Lorealis & 175\\
        @Download-Links & 126\\
        @TautaraHack & 62 \\
        @redirect-5f2cx25f & 61\\
        @pengwincheat & 50 \\\hline
    \end{tabular}}
    \label{tab:num_youtube_channel}
\end{table}

The potential for familial impact further amplifies this threat. A recent survey~\cite{Lucchesi2020-ty} found that 35\% of households share a single PC among family members, compared to only 10\% for smartphones. This high rate of shared device usage means that an accidental malware infection of one child could easily become a family-wide problem, underscoring the urgent need for effective countermeasures against MalTube attacks.

Our research confirmed existing knowledge of MalTube tactics and uncovered new insights into the sophistication of these attacks.

\begin{itemize}[leftmargin=0.2in]
    \item \textbf{SEO Strategies and Multilingual Approach:} We confirmed previous findings regarding the attackers' use of search engine optimization (SEO) techniques. These include embedding topic-related keywords in video descriptions to improve searchability and using multiple languages to expand their target audience.
    
    \item \textbf{Engaging Thumbnails:} We identified the strategic use of character images and screenshots demonstrating the alleged effects of the illegal software in video thumbnails. While some thumbnails were reused across multiple videos, we noted a wide variety of unique designs, all designed to attract user attention.
    
    \item \textbf{User-friendly download instructions:} The attackers demonstrated a clear focus on guiding users through the download process with minimal confusion. Detailed instructions were consistently repeated in thumbnails, video content, descriptions, and intermediary sites. Providing passwords for archived malware further streamlined the infection process for targeted users.
    
    \item \textbf{Novel use of platform features:} Our research uncovered the innovative use of YouTube channel description sections as intermediary sites within the MalTube attack infrastructure. Some of these channels even used custom character logos, creating a sense of brand identity to further engage potential victims.
    
    \item \textbf{Simplified Click Chain:} We observed that MalTube attacks typically involve fewer intermediate steps than other malware distribution methods. This streamlined approach, necessitated by the reliance on user-initiated downloads, results in more direct paths to malware acquisition.
\end{itemize}

Combined, these characteristics demonstrate that MalTube attacks prioritize user engagement and minimizing confusion or abandonment during the manual malware download process. The use of leet speak and emoticons in passwords suggests an attempt to balance human readability with evasion of automated detection systems.

Although the reliance on user action represents a potential weakness in the attack chain, the sophisticated social engineering employed by the MalTube operators makes this a serious threat. The amount of effort the attackers put into guiding users through the infection process underscores the potential effectiveness of this distribution method.

\section{Discussion}\label{sec:discussion}

Here, we build on the results of our measurement study to discuss potential countermeasures against MalTube attacks, address the ethical considerations of our research, and acknowledge the limitations of our approach.

\subsection{Countermeasures}\label{subsec:Countermeasure}

\subsubsection{Detecting MalTube videos}

Our observations indicate that a significant portion of MalTube videos persist over time. Of the 14,363 videos monitored between December 20, 2023 and April 30, 2024, 9,155 remained accessible throughout the study period. This persistence highlights the need for more efficient detection and removal processes.

We propose using the VIPER system to continually expand and refine search terms to enable more comprehensive and timely identification of MalTube content. Although we expect attack methods to evolve, certain invariant characteristics of MalTube videos are likely to persist. In particular, the need to communicate download instructions and passwords to users is critical to attack success and difficult for perpetrators to significantly alter. We believe these consistent elements will remain valuable for future detection efforts.

\subsubsection{Detecting intermediary and malware distribution sites}

The nature of MalTube attacks, which rely on user-initiated downloads rather than automated redirects, shapes the characteristics of the associated intermediary and distribution sites. Unlike drive-by download attacks~\cite{Sood2016-ie}, MalTube sites must prioritize clear instructions and maintain user engagement throughout the download process. These requirements create distinct patterns that can be used for detection.

Although there are some unique distribution sites, our research revealed a limited number of commonly used intermediaries and download platforms (e.g., Mediafire). This concentration provides an opportunity for targeted monitoring and potential access restrictions to these high-risk sites.
Even for custom distribution sites, we have identified common characteristics that can aid in detection. These include the use of specific terminology (e.g., ``cheat'') to appeal to users, and instructions for circumventing security measures such as password prompts.

\subsubsection{User intervention and education}

The reliance of MalTube attacks on user action suggests that raising awareness and educating potential victims could significantly reduce infection rates. Unlike drive-by download attacks that exploit technical vulnerabilities, MalTube attacks primarily use social engineering tactics. This characteristic presents both a challenge and an opportunity for defense strategies.

We believe that the ecosystem model developed through this study can serve as a valuable educational tool to help users understand the mechanics and risks associated with these attacks. By visualizing the entire attack chain, from initial video encounter to malware execution, users can better understand the multiple decision points at which they can interrupt the attack process. This model could be adapted into interactive simulations or gamified learning experiences, particularly tailored for younger users who are most at risk.

Previous research on phishing attacks~\cite{Franz2021-ir} has highlighted the importance of timely intervention when users encounter potential threats. This principle also applies to MalTube attacks. Developing effective warning and education systems that can be integrated into browsers, YouTube applications, and other potential points of contact with MalTube content remains an important area for future work. These systems could leverage machine learning algorithms trained on our dataset to identify potential MalTube videos in real time and provide contextual warnings to users.

In addition, our findings suggest that a multi-pronged educational approach is needed. This should include:

\begin{itemize}
    \item Age-appropriate cybersecurity curricula in schools, focusing on the risks of downloading unauthorized software and game cheats.
    \item Parent education programs to help adults understand and communicate with their children about online risks.
    \item Collaboration with gaming communities and influencers to raise awareness of the tactics of MalTube among their audiences.
    \item Platform-specific guidelines and pop-up reminders about safe browsing practices, specifically when users search for terms commonly associated with MalTube content.
\end{itemize}

\subsection{Ethical Considerations}\label{subsec:Ethics}

Throughout our research, we maintained a strong commitment to ethical practices and responsible disclosure. Our approach to data collection and analysis was guided by several key principles. We carefully timed our use of the YouTube API to avoid placing an undue burden on the infrastructure of the platform and to ensure that our research activities did not interfere with normal platform operations or user experience. We are actively engaged in the process of sharing our findings regarding the characteristics of MalTube with the security operation team of YouTube, conducting this disclosure in accordance with best practices for vulnerability reporting and allowing the platform sufficient time to assess and address the threats identified.

Throughout our data collection and analysis processes, we prioritized user privacy and ensured that no personally identifiable information was collected or stored as part of our research. Although our study involved identifying malware distribution channels, we took great care not to inadvertently spread malicious software by conducting our analysis in isolated, secure environments to avoid any risk of infection or further distribution.

\subsection{Limitations and Future Work}\label{subsec:limitation}

Although our approach is comprehensive, it is subject to several limitations that may affect the generalizability and completeness of our findings. Here, we discuss these limitations and suggest future work to address them.

\subsubsection{Limitations of keyword-based detection}

The keyword-based detection method used by VIPER may not capture all the activities of MalTube. Although we have made efforts to expand our keyword set based on detected video tags and information from malware distribution sites, it remains a challenge to keep up with the rapidly evolving terminology used by attackers.

\noindent\textbf{Future work:} To address this limitation, future research could explore the use of more advanced natural language processing techniques, including deep learning models trained on a diverse corpus of MalTube content. Additionally, implementing a semi-supervised learning approach could help identify new keywords and phrases associated with MalTube attacks in real time.

\subsubsection{Incomplete verification of malware distribution}

Although we have confirmed malware downloads from a subset of identified sites, fully automating this process has proven difficult. Concerns about alerting attackers to our monitoring activities and potentially triggering access controls limited our ability to verify all identified distribution points. Therefore, our dataset may include some videos that do not ultimately result in malware downloads.

\noindent\textbf{Future work:} Future studies could develop more sophisticated automated verification systems that more closely mimic human behavior to avoid detection. Collaboration with cybersecurity firms or law enforcement agencies could also provide safer means of verifying malware distribution without compromising the integrity of the research or inadvertently aiding attackers.

\subsubsection{Potential inclusion of non-malicious content}

To mitigate the limitation of potentially including non-malicious content, we manually reviewed a sample of videos to identify characteristics of benign content (e.g., links to legitimate social media profiles) and incorporated these findings into our filtering process. Although this approach helps reduce false positives, it cannot completely eliminate the possibility of some non-malicious content in our dataset.

\noindent\textbf{Future work:} Future research could focus on developing more robust automated classification systems using machine learning techniques. This could include creating a larger, manually verified dataset for training and incorporating multimodal analysis that considers video content, audio, and metadata. In addition, implementing a continuous feedback loop with platform moderators could help refine classification accuracy over time.

\section{Related Work}\label{sec:relatedwork}

This section reviews existing literature relevant to our study, focusing on research on malicious activities on video-sharing platforms, detection of spam and malicious links, and various forms of user deception in online environments.

\subsection{Malicious Activities on Video Sharing Platforms}

Several studies have examined criminal or undesirable behavior on YouTube and similar platforms. Chu et al. analyzed forum posts and YouTube account trading sites to uncover illegitimate monetization methods~\cite{Chu_undated-kk}. Nasir et al. examined YouTube videos promoting fraudulent money-stealing apps and found that these scams primarily target users in developing regions~\cite{Nasir2022-gz}. Heuer et al. investigated potential political bias in the recommendation system of YouTube in Germany by tracing chains of recommended videos~\cite{Heuer2021-cc}.

Our research differs from these studies by focusing on a specific type of malicious content in which users inadvertently participate in illegal activities. This user complicity potentially complicates reporting and removal processes, necessitating our investigation.

\subsection{Detecting Spam and Malicious Links}

Researchers have proposed various methods for detecting spam comments and malicious links on video-sharing platforms. O-Callaghan et al. developed a network profiling approach to identify spam comments on YouTube~\cite{O-Callaghan2021-rw}, whereas Valpadasu et al. proposed a Naive Bayes classification algorithm for the same purpose~\cite{Valpadasu2023-bu}. Alshamrani et al. detected attacks in YouTube comments that attempt to trick children into opening suspicious URLs~\cite{Alshamrani2020-mu}. In a related study, Janet et al. proposed a method for detecting similar attacks in Twitch chat rooms~\cite{Janet2022-ba}.

Our study goes beyond these approaches by examining the malicious links, as well as the videos and accounts used to distribute them. Additionally, we determined sophisticated evasion techniques used by the attackers, such as the use of intermediary sites that host pages from other YouTube accounts.

\subsection{User Deception in Software Downloads and Scams}

Research on tactics used to trick users into downloading software or falling for financial scams has also been conducted. Badawi et al. examined game hack scams that trick users into providing personal information or downloading suspicious files in exchange for in-game resources~\cite{Badawi2020-sn}. Koide et al. investigated methods of disguising malware as antivirus software~\cite{Koide2021-ar}, whereas Rauti et al. provided detailed insights into technical support scams by deliberately falling victim to real fraudsters~\cite{Rauti2017-ge}. Vardhan et al. investigated scams perpetrated through direct messages on social media platforms~\cite{Vardhan2023-gz}.

Although these studies provide valuable insights into various forms of user deception, our research differs in its focus on the unique ecosystem of video-based malware distribution. Unlike conventional scams or fraudulent downloads, MalTube attacks leverage the credibility and reach of popular video-sharing platforms to distribute malware. Our study examines the deception tactics used and the complex infrastructure supporting these attacks, including the use of intermediary sites and the exploitation of platform-specific features. In addition, we explore the psychological aspects unique to this attack vector, such as exploiting the desire of users for free premium software and game cheats that can lower their guard against potential threats.

\subsection{Email-Based Phishing and Malware Distribution}

Email-based phishing and malware distribution continue to be persistent threats~\cite{Oest2020-nu,Ho2017-qj,Hu2018-vi,Simoiu2020-of}. Researchers have developed various countermeasures, including detection methods for suspicious emails and phishing sites~\cite{Liang2016-wr,Duman2016-cf,Cidon2019-ra}, and user support techniques such as alerts~\cite{Stembert2015-nc, Petelka2019-lg,UnknownUnknown-oi}, improved email client user interfaces~\cite{Marforio2016-cr,Nicholson2017-pk}, and security training programs~\cite{Daniele-Lain-Kari-Kostiainen-and-Srdjan-CapkunUnknown-oc,Carella2017-cu,UnknownUnknown-oi}.

Although these studies do not directly address MalTube attacks, they highlight the importance of observing and developing countermeasures against user deception tactics in different online contexts. Our research builds on this foundation and extends the understanding of such threats to the realm of video-based malware distribution.
\section{Conclusion}\label{sec:conclusion}

This study presents a comprehensive analysis of MalTube, an emerging malware distribution technique that exploits video sharing platforms, and addresses a critical gap in the existing literature by systematically examining the attack ecosystem, attacker strategies, and user deception methods. We designed and implemented VIPER, an advanced monitoring system that collected and analyzed data from 14,363 malicious videos over a four-month period, revealing that MalTube attacks target a wide range of users, particularly gamers and minors. Our analysis revealed sophisticated attacker channel management strategies and meticulous social engineering tactics. We identified a key weakness in the MalTube attack methodology: the need for attackers to provide explicit download instructions and passwords in their videos, which can serve as a distinguishing feature for detection. We discovered 1,269 unique FQDNs associated with MalTube attacks and revealed an extensive supporting infrastructure, including the novel use of YouTube channel descriptions as intermediary sites. These findings provide a foundation for developing effective countermeasures, highlighting the need for improved detection mechanisms, user education programs, platform collaboration, and infrastructure disruption strategies. Our research contributes to the literature on cyber threats and provides insights for security professionals, platform operators, and policy makers. As online platforms continue to evolve, the results of this study can inform the development of more robust security measures and user protection strategies. Further research in this area could potentially lead to improved methods for detecting and mitigating similar socially engineered attacks in dynamic online environments.

\bibliographystyle{IEEEtranS}
\bibliography{reference}

\end{document}